\documentstyle[aps]{revtex}
\begin{document}
\title{Black-hole radiation, the fundamental area unit, and the
  spectrum of particle species} 
\author{Shahar Hod}
\address{The Racah Institute for Physics, The
Hebrew University, Jerusalem 91904, Israel}
\date{\today}
\maketitle

\begin{abstract}

Bekenstein and Mukhanov have put forward the idea that, in a quantum
theory of gravity a black hole should have a discrete mass spectrum with
concomitant {\it discrete} line emission. We note that a direct 
consequence of this intriguing prediction is that compared with blackbody radiation, black-hole
radiance is {\it less} entropic. We calculate the ratio of entropy emission
rate from a quantum black hole to the rate of black-hole
entropy decrease, a quantity which according to the generalized second
law (GSL) of thermodynamics should be larger than unity. Implications
of our results for the GSL, for the value of the fundamental area
unit in quantum gravity, and for the spectrum of massless particles in
nature are discussed.

\end{abstract}
\bigskip

One of the most remarkable theoretical predictions of the former century is
Hawking's celebrated result that black holes are {\it quantum}
mechanically unstable: they decay by the emission of a (filtered)
thermal radiation \cite{Haw1}. According to Hawking's result, the
black hole emits quanta of {\it all} frequencies, distributed
according to the usual black-body spectrum (with a gray-body factor
which represents the imprint of passage through the curvature
potential surrounding the black hole).

However, Hawking's prediction of black-hole evaporation is at a
semiclassical level in the sense that the matter fields are treated
quantum mechanically, but the spacetime (and the black hole itself)
are treated {\it classically}. One therefore suspects some
modifications of the character of the radiation when quantum
properties of the {\it black hole itself} are properly taken into account.

The quantization of black holes was proposed long ago by Bekenstein
\cite{Beken1}. Based on the remarkable observation that the horizon
area of a nonextremal black hole behaves as a classical 
adiabatic invariant, and in the spirit of Ehrenfest principle
\cite{Ehren}, any classical adiabatic invariant
corresponds to a quantum entity with {\it discrete} spectrum,
Bekenstein \cite{Beken1} conjectured that the horizon area of a quantum
black hole should have a discrete eigenvalue spectrum of the form

\begin{equation}\label{Eq1}
A_n=\gamma {\ell^2_P} \cdot n\ \ \ ;\ \ \ n=1,2,\ldots\ \  ,
\end{equation}
where $\gamma$ is a dimensionless constant, and 
$\ell_P=\left({G \over {c^3}}\right)^{1/2} {\hbar}^{1/2}$ is the
Planck length (we use gravitational units in which $G=c=1$). This type
of quantization-law has since been revived on various grounds
\cite{Muk,BekMuk,Beken2,Maz,Ko,Mag,Lo,Pe,LoMa,BaKu,Ka1,Mak,Hod1,Hod2,Hod3,VaWi,Ka2,BoKa,Ah,Ga}
(most of these derivations have been made in the last few years). 
In particular, Mukhanov and Bekenstein
\cite{Muk,BekMuk,Beken2} used a combination of thermodynamic 
(the area-entropy relation $S_{BH}=A/4\hbar$ for
black holes) and statistical physics (the Boltzmann-Einstein
formula) arguments, and found that the dimensionless constant $\gamma$ in
Eq. (\ref{Eq1}) should be of the form $\gamma =4 \ln \beta$, with $\beta=2,
3, \ldots$ (this corresponds to a degeneracy factor of $\beta^n$ for the
$n$th area level). Using Bohr's correspondence principle Hod \cite{Hod1}
has recently given evidence in favor of the value $\beta=3$.

The discrete mass (area) spectrum implies
a {\it discrete} line emission from a quantum black hole; the
radiation emitted by the black hole will be at integer multiples of
the fundamental frequency $\omega_0=\ln \beta/8\pi M$ \cite{BekMuk}. 

We note that a direct 
consequence of the {\it discrete} spectrum is that compared with
blackbody radiation, black-hole
radiance is {\it less} entropic. In fact, one expects the entropy
of the radiation emitted by the black hole (or equivalently, the total number of
configurations evaporated by the black hole) to {\it decrease} as
$\beta$ increases: The entropy of the radiation should be maximal when
the various transitions have equal probabilities, but the fundamental
transition $n \to n-1$ becomes more and more dominant as the value of
$\beta$ increases. 

The entropy of a system measures one's {\it lack of information} about
its actual internal configuration \cite{Shan,Jayn,Beken3}. 
Suppose that all that is known about the system's internal configuration is
that it may be found in any of a number of states, with probability
$p_n$ for the $n$th state. Then the entropy associated with the system
is given by Shannon's well-known relation $S=-\sum_{n} p_n \ln p_n$. 

The probability for a black hole to emit a specific quantum 
should be proportional to the 
degeneracy of the {\it final} black-hole quantum state, to the gray-body
factor $\Gamma$ (representing a scattering of the quantum 
off the spacetime curvature surrounding the black hole), and to
the square of the matrix element. In the spirit of the original
treatment of Bekenstein and Mukhanov \cite{BekMuk} 
we assume that the matrix element does not vary much as
one goes from a nearest neighbor transition to one between somewhat
farther neighbors. Thus, aside from an overall normalization 
factor, the matrix element does not enter into our 
simple estimate. This assumption is further supported by a recent
analysis of Massar and Parentani \cite{MaPa}. Thus, the probability
$p_k$ to jump $k$ steps in the mass (area) ladder is proportional to
$\Gamma(k) \beta^{-k}$.

{\it A toy model.} 
To gain some insight into the physical problem, we shall first
consider a simplified toy model. It is well known that for massless fields,
$\Gamma(M \omega)$ approaches $0$ in the low-frequency limit, and has a
high-frequency limit of $1$ \cite{Page}. A rough approximation of this
effect can be achieved by introducing a low frequency cutoff at some
$\omega=\omega_c$ \cite{Zur}. That is, $\Gamma(\varpi)=0$ for
$\varpi < \varpi_c$ and $\Gamma(\varpi)=1$ otherwise, where
$\varpi \equiv M\omega$.  

The ratio $R=|\dot S_{rad} / \dot S_{BH}|$ of entropy emission
rate from the quantum black hole to the rate of black-hole
entropy decrease is given by

\begin{equation}\label{Eq2}
R={{-{\sum_{i=1}^{N_s}}{\sum_{k=1}^{\infty}}C\Gamma(k)\beta^{-k} \ln
  [C\Gamma(k)\beta^{-k}]} \over {{\sum_{i=1}^{N_s}}{\sum_{k=1
}^{\infty}}C \Gamma(k)\beta^{-k} k \ln\beta}}\  ,
\end{equation}
where $N_s$ is the effective number
of (massless) particle species emitted ($N_s$ takes into account the
various modes emitted), and $C$ is a normalization factor, 
defined by the normalization condition

\begin{equation}\label{Eq3}
{\sum_{i=1}^{N_s}}{\sum_{k=1}^{\infty}}C\Gamma(k)\beta^{-k}=1\  .
\end{equation}
Equation (\ref{Eq2}) for $R$ yields

\begin{equation}\label{Eq4}
R=1-{{\ln[(1-\beta^{-1})\beta^{k_0}/N_s]} 
\over {\ln \beta}} {{\beta -1} \over {1+k_0(\beta-1)}}\  , 
\end{equation}
where $k_0 \equiv \omega_c / \omega_0$.
As expected, the number of configurations evaporated by the black hole
(or equivalently, the value of $R$) increases with increasing number $N_s$ of massless particles in nature. 

For the emission process to respect the GSL, $R$ should be larger than
(or equal) unity. Thus, one finds that the number $N_s$ of massless
particles should satisfy the relation

\begin{equation}\label{Eq5}
(1-\beta^{-1})\exp(8\pi \varpi_c) \leq N_s\  .
\end{equation}
In accordance with our previous expectation, 
the entropy of the radiation emitted by the
black hole decreases as the value of $\beta$ increases. Thus, the
minimal value allowed for $N_s$ increases with increasing value of $\beta$.

One can {\it estimate} the maximally 
allowed value of $\beta$ (which is consistent with the GSL) by taking
$\varpi_c \simeq 0.2$ (the location of the peak in the total power 
spectrum \cite{Page}), and $N_s \simeq 112$
(this estimate takes into account the modes which make the 
dominant contribution to the black-hole spectrum \cite{Page}: three species of neutrinos with modes having
$j=1/2,3/2,5/2$, a photon with modes having $j=1,2$, and a graviton with
modes having $j=2,3$). 
Taking cognizance of Eq. (\ref{Eq5}) one finds $\beta_{max} \simeq
3.7$ (which in practice implies a maximal value of $3$ for $\beta$). An exact
calculation to be given below reveals that this is a fairly good estimate.

{\it An exact calculation.} The most significant correction to our
toy-model will result from the mode-dependent of the gray-body 
factor $\Gamma_{s,\varpi,j,m}$ (first calculated numerically by Page
\cite{Page}), which is a function of the particle
species $s$, the frequency $\varpi$, and the angular-momentum
$(j,m)$.

We consider the emission of a canonical set of three
species of neutrinos, photon, and a graviton. 
The ratio $R=|\dot S_{rad} / \dot S_{BH}|$ of entropy emission
rate from the quantum black hole to the rate of black-hole
entropy decrease is given in Table \ref{Tab1}. For the evaporation
process to respect the GSL, $R$ should be larger (or equal) than
unity. We therefore conclude that the maximally allowed value for $\beta$
is $3$ (see Table \ref{Tab1}).

{\it Summary and physical implications.} 
We have studied the emission of radiation from a {\it quantized} black hole,
which is characterized by a {\it discrete} mass spectrum. [The discrete
mass spectrum (with an evenly spaced area levels) has been derived on 
various grounds (see e.g.,
\cite{Beken1,Muk,BekMuk,Beken2,Maz,Ko,Mag,Lo,Pe,LoMa,BaKu,Ka1,Mak,Hod1,Hod2,Hod3,VaWi,Ka2,BoKa,Ah,Ga})].
In particular, we have considered the quantum evaporation process from the point of
view of the generalized second law of thermodynamics. It was
conjectured, and demonstrated explicitly by a simplified toy-model, that the entropy
emitted from the black hole decreases as the area spacing
increases. This allowed us to derive an upper bound to the value of
the fundamental area spacing $\Delta A=4\ell^2_P \ln 3$. Remarkably, this
value agrees with the one recently derived based on Bohr's correspondence
principle \cite{Hod1}.

We note, however, that in recent years there is a growing evidence
that neutrinos have {\it finite} masses. This implies that black holes
with masses larger than $\hbar/m_{\nu}$ will not emit neutrinos with
mass $m_{\nu}$ in a significant rate. As a direct consequence, the
entropy emitted from the black hole would {\it decrease} because the 
aveliable phase space becomes smaller. In fact, we learn form Table
\ref{Tab1} that if there were only two kinds of massless neutrinos
species in nature, then the largest value allowed for $\beta$ (which
is still consistent with the GSL) is $2$. (The GSL would be violated if there was no more than one
massless neutrino specie.)

In view of the previous discussion, we see only few possible solutions to the
apparent violation of the GSL:

(1) The uniformally spaced area spectrum, 
first suggested by Bekenstein \cite{Beken1} (and rederived on various
different grounds by many authors \cite{Muk,BekMuk,Beken2,Maz,Ko,Mag,Lo,Pe,LoMa,BaKu,Ka1,Mak,Hod1,Hod2,Hod3,VaWi,Ka2,BoKa,Ah,Ga}),
does not hold in reality.

(2) The quantum matrix elements vary considerably as
one goes from a nearest neighbor transition to one between farther
neighbors. (We recall that we assumed in the spirit of the original
analysis of Bekenstein and Mukhanov \cite{BekMuk}, and in accordance
with the recent analysis of Massar and Parentani \cite{MaPa}, that 
the quantum matrix element does not vary much as
one goes from a nearest neighbor transition to one between farther
neighbors.) 

(3) Neutrinos are massless after all. 
In this case we have shown that the GSL is respected during the
evaporation process (provided $\beta \leq 3$).

(4) There exists a yet-unknown massless particle(s) in nature
apart from the photon and the graviton. This would increase the 
number of configurations (or equivalently, the entropy) evaporated by
the black hole. 

Evidently, this solution 
to the apparent violation of the GSL is much more `exotic' than the
former ones. However, one should {\it not} rule it out; In fact
W. Pauli \cite{Pau} proposed the existence of the neutrino itself as a
``desperate'' solution in order to prevent a violation of the law of the 
conservation of energy during the process of beta-decay. Here we 
conjecture the existence of a yet-unknown massless particle(s) in order
to prevent a violation of another sacred law of physics, the
(generalized) second law. 
The conjectured particle may therefore be
appropriately named the `entropon'.

\bigskip
\noindent
{\bf ACKNOWLEDGMENTS}
\bigskip

I thank Jacob D. Bekenstein for helpful discussions.
This research was supported by a grant from the Israel Science Foundation.

\begin{table}
\caption{The ratio $R=|\dot S_{rad} / \dot S_{BH}|$ of entropy emission
rate from a quantum black hole to the rate of black-hole entropy decrease.}
\label{Tab1}
\begin{tabular}{lll}
$\beta$ &$N \nu =2$ & $N \nu =3$\\
\tableline
$2$ & $1.042$ & $1.119$ \\
$3$ & $0.940$ & $1.016$ \\
$4$ & $0.888$ & $0.962$  \\
\end{tabular}
\end{table}


\begin{thebibliography}{99}

\bibitem{Haw1} S. W. Hawking, Commun. Math. Phys. {\bf 43}, 199 (1975).

\bibitem{Beken1} J. D. Bekenstein, Lett. Nuovo Cimento {\bf 11}, 467 (1974).

\bibitem{Ehren} See for example M. Born, Atomic Physics (Blackie,
  London, 1969), eighth edition.

\bibitem{Muk} V. Mukhanov, JETP Lett. {\bf 44}, 63 (1986).

\bibitem{BekMuk} J. D. Bekenstein and V. F. Mukhanov, Phys. Lett. B
  {\bf 360}, 7 (1995).

\bibitem{Beken2} J. D. Bekenstein in XVII Brazilian National Meeting
  on Particles and Fields, eds. A. J. da Silva et. al. (Brazilian
  Physical Society, Sao Paulo, 1996), J. D. Bekenstein in Proceedings of the VIII
  Marcel Grossmann Meeting on General Relativity, eds. T. Piran and
  R. Ruffini (World Scientific , Singapore, 1998).

\bibitem{Maz} P. O. Mazur, Phys. Rev. Lett. {\bf 57}, 929 (1987).

\bibitem{Ko} Ya. I. Kogan, JETP Lett. {\bf 44}, 267 (1986).

\bibitem{Mag} M. Maggiore, Nucl. Phys. B {\bf 429}, 205 (1994).

\bibitem{Lo} C. O. Lousto, Phys. Rev. D {\bf 51}, 1733 (1995).

\bibitem{Pe} Y. Peleg, Phys. Lett. B {\bf 356}, 462 (1995).

\bibitem{LoMa} J. Louko and J. M\"akel\"a, Phys. Rev. D {\bf
    54}, 4982 (1996).

\bibitem{BaKu} A. Barvinsky and G. Kunstatter, Phys. Lett. B {\bf
    389}, 231 (1996).

\bibitem{Ka1} H. A. Kastrup, Phys. Lett. B {\bf 385}, 75 (1996).

\bibitem{Mak} J. M\"akel\"a, Phys. Lett. B {\bf 390}, 115 (1997).

\bibitem{Hod1} S. Hod, Phys. Rev. Lett. {\bf 81}, 4293 (1998).

\bibitem{Hod2} S. Hod, Phys. Rev. D {\bf 59}, 024014 (1999).

\bibitem{Hod3} S. Hod, Gen. Rel. Grav. {\bf 31}, 1639 (1999).

\bibitem{VaWi} C. Vaz and L. Witten, Phys.Rev. D {\bf 60}, 024009 (1999).

\bibitem{Ka2} H. A. Kastrup, e-print gr-qc/9906104.

\bibitem{BoKa} M. Bojowald and H. A. Kastrup, e-print hep-th/9907043.

\bibitem{Ah} D. V. Ahluwalia, Int. J. Mod. Phys. D {\bf 8}, 651 (1999).

\bibitem{Ga} R. Garattini, e-print gr-qc/9910037; e-print gr-qc/0003090.

\bibitem{Shan} C. E. Shannon and W. Weaver, 
{\it The Mathematical Theory of Communications} (University of Illinois Press, Urbana, 1949).

\bibitem{Jayn} E. T. Jaynes, Phys. Rev. {\bf 106}, 620 (1957); {\bf 108}, 171 (1957).

\bibitem{Beken3} J. D. Bekenstein, Phys. Rev. D {\bf 7}, 2333 (1973).

\bibitem{MaPa} S. Massar and R. Parentani, 
e-print gr-qc/9903027, Nucl. Phys. B (to be published).

\bibitem{Page} D. N. Page, Phys. Rev. D {\bf 13}, 198 (1976);
  Phys. Rev. D {\bf 14}, 3260 (1976); Phys. Rev. D {\bf 16}, 2402 (1977).

\bibitem{Zur} W. H. Zurek, Phys. Rev. Lett. {\bf 49}, 1683 (1982). 

\bibitem{Pau} W. Pauli, Phys. Rev. {\bf 38}, 579 (1931).

\end{thebibliography}
\end{document}